\documentclass{PoS}
\usepackage{bm}
\title{The role of angular ordering condition in Parton Branching transverse momentum dependent (TMD) PDFs and DY transverse momentum spectrum at LHC}

\ShortTitle{Angular ordering condition in Parton Branching TMDs}

\author{\speaker{Aleksandra Lelek}\\
        UAntwerp\\
        E-mail: \email{aleksandra.lelek@uantwerpen.be}}

\abstract{We discuss the parton branching (PB) evolution equation for transverse momentum dependent (TMD) parton distribution functions (PDFs), especially we concentrate on the angular ordering constrain and its effect on the TMD distributions.

We discuss application of the PB TMDs to precise prediction of Drell-Yan transverse momentum spectrum at LHC, stressing the role of angular ordering in the low transverse momentum region.

We compare the PB implementation of the angular ordering condition with the Kimber-Martin-Ryskin-Watt (KMRW), both at the analytical and numerical level. }

\FullConference{
European Physical Society Conference on High Energy Physics - EPS-HEP2019 -\\
			10-17 July, 2019\\
			Ghent, Belgium}

\begin{document}

\section{Why TMDs?}

Collinear factorization theorem \cite{Collins:1989gx} is a successful tool commonly used to obtain QCD predictions for sufficiently inclusive collider observables with 
only one hard scale relevant. However, for observables with a second scale involved (as e.g. transverse momentum in the final state) the collinear approximation is  not sufficient and  transverse momentum dependent (TMD) factorization theorems should be used to obtain precise predictions \cite{Angeles-Martinez:2015sea}.\\
 
One of the open problems  is a consistent kinematics treatment in parton showers (PS). As reported in \cite{Hautmann:2012dw},  when the PS is added on top of the collinear matrix element (ME), the energy-momentum conservation requires that the kinematics in the ME must be manipulated to compensate for the transverse momentum generated in the PS procedure. This leads to a mismatch between the 4 momenta of the partons in the ME and the parton distribution functions (PDFs) initially used for the generation. 
  Based on the TMD factorization \cite{Catani:1990xk}, an approach was proposed \cite{Jung:2000hk} where off-shell (TMD) ME could be used together with TMD PDFs (abbreviated here as TMDs). Higher order QCD corrections could be included via TMD PS. Thanks to this procedure, the transverse momentum generated in the ME would not be changed by adding the PS and the kinematics would be properly treated. The key ingredient to make this approach possible is the availability of the TMDs. \\

The Parton Branching (PB) method developed in the last few years \cite{Hautmann:2017xtx,Hautmann:2017fcj} is an approach which allows one to obtain TMDs. 
The method incorporates the phenomena of angular ordering 
\cite{Marchesini:1983bm,Marchesini:1987cf}
 - the fact that the colour coherence effects result in certain ordering in angle of the soft gluons emitted in QCD radiation.  
In the following we examine the role of the angular ordering condition in obtaining precise QCD predictions for LHC observables, concentrating especially on the scale dependent soft gluon resolution parameter which was extensively studied in  \cite{Hautmann:2019biw}.

\section{Angular ordering in PB}

The PB TMD evolution equation with angular ordering has the following form
~\cite{Hautmann:2017fcj}  
\begin{eqnarray}
&& \widetilde{A}_a\left( x, {\bm k}, \mu^2\right) = \Delta_a\left(\mu^2, \mu_0^{2}\right)\widetilde{A}_a\left( x, {\bm k}, \mu_0^2\right)+ 
 \sum_b\int \frac{\textrm{d}^2{\boldsymbol \mu}^{\prime}}{\pi {\mu}^{\prime 2}}\Theta\left(\mu^{2}-\mu^{\prime 2}\right)\Theta\left(\mu^{\prime 2}-\mu_0^{ 2}\right)
 \nonumber \\ 
 &\times& 
  \int_x^{1-\frac{q_0}{\mu^{\prime}}}\textrm{d}z \
  {  { \Delta_a\left(\mu^2, \mu_0^2  \right)  } \over 
  { \Delta_a\left(\mu^{\prime 2}, \mu_0^2 \right) } } \ 
  P_{ab}^{R}\left(z,\alpha_s((1-z)^2\mu^{\prime 2})\right)\widetilde{A}_b\left( \frac{x}{z},  {\bm k} + (1-z){\boldsymbol \mu}^\prime, \mu^{\prime 2}\right) \; , 
\label{eq:tmdevol}
\end{eqnarray} 
where $\widetilde{A}_a\left( x, {\bm k}, \mu^2\right)= x A_a\left( x, {\bm k}, \mu^2\right)$ is the momentum-weighted 
TMD distribution of a parton of flavor $a$, carrying the longitudinal momentum  fraction $x$ of the proton's momentum and  transverse momentum ${\bm k}$\footnote{The following    notation is used: $k=(k^0, k^1, k^2, k^3)=(E_{k}, {\bm k}, k^3)$, where ${\bm k}=(k^1, k^2)$, and $k_{\bot} = | {\bm k} | $. } 
at the evolution scale $\mu$;  $z$ and ${\boldsymbol \mu}^\prime$ are the branching variables, with $z$ being the ratio of the proton's longitudinal momentum fraction carried by parton $a$ and $b$, and  $ \mu^\prime = \sqrt{ {\boldsymbol \mu}^{\prime 2}}$ the  scale at which the branching happens. 
 $P_{ab}^R$ are the real-emission splitting functions and 
 $\Delta_a$ is   the Sudakov form factor,  given by  
$\Delta_a(\mu^2, \mu_0^2)  = \exp\left[-\sum_b \int_{\mu_0^2}^{\mu^2}\frac{\textrm{d}\mu^{\prime 2}}{\mu^{\prime 2}} \int_0^{1-q_0/\mu^{\prime}}\textrm{d}z \  
\ z \ P_{ba}^{R}\left(z,\alpha_s\left((1-z)^2\mu^{\prime 2}\right)\right) \right]$.
  The initial  
evolution scale is denoted by $\mu_0$. The parameter $q_0$ defining the soft gluon resolution scale, is the minimum transverse momentum of the emitted parton, with which it can be resolved.  The {\it{dynamical}} resolution scale $z_M \equiv 1-\frac{q_0}{\mu^{\prime}}$ depends on the scale of the branching $\mu^{\prime}$. Eq.~(\ref{eq:tmdevol}) defines also the relation between the transverse momentum of the emitted parton  and the scale at which the branching happens $q_{\bot} = (1-z)\mu^{\prime}$. With that one can see that the scale at which the $\alpha_s$ is evaluated, is the transverse momentum of the emitted parton. The transverse momentum ${\bm k}$ is calculated by 
 combining the intrinsic transverse momentum with  the transverse momenta 
 emitted in all branchings.

In \cite{Martinez:2018jxt}  the fit of PB integrated TMDs to precision measurements of deep inelastic scattering (DIS) cross sections at HERA was performed using \texttt{xFitter} \cite{Alekhin:2014irh}. TMDs obtained this way 
 were used later to obtain predictions for Z boson $p_{\bot}$ spectrum \cite{Martinez:2018jxt, Martinez:2019mwt}. Good agreement with the ATLAS data \cite{Aad:2015auj} was obtained. The results in \cite{Martinez:2018jxt, Martinez:2019mwt}  were obtained with fixed value of $z_M$. In the next section we will investigate the influence of the dynamical $z_M$ on the results.

\section{iTMDs with angular ordering}

Collinear distributions (reffered to also as the integrated TMDs, iTMDs) can be obtained from  eq.~(\ref{eq:tmdevol}) by integrating the TMD distributions over the transverse momenta \footnote{ As shown in ~\cite{Hautmann:2017xtx,Hautmann:2017fcj}, for  $z_M \to 1$ and $\alpha_s \to \alpha_s (\mu^{\prime 2})$  these are the  collinear parton distribution functions satisfying  
DGLAP evolution equations ~\cite{Gribov:1972ri}.}
\begin{equation} 
 \widetilde{f}_a\left(x, \mu^2\right) =  \int { {\textrm{d}^2{\bm k}  } \over \pi } \ \widetilde{A}_a\left(x, {\bm k}, \mu^2\right) \;.
\label{integratingAoverKt}
\end{equation}
With angular ordering condition,  the evolution equation for $ \widetilde{f}_a $ 
is given  by 
\begin{eqnarray}
\label{PBangular}
  &&\widetilde{f}_{a}(x,\mu ^{2}) = 
  \Delta_a(\mu ^{2} , \mu_0^2)    
  \widetilde{f}_{a}(x,\mu _{0} ^{2})  + 
  \sum _{b}  \int_{\mu_{0} ^{2}}^{ \mu ^{2}}  \frac{d \mu ^{\prime 2}}{\mu ^{\prime 2}}     \int _{x}^{1}  dz  
   \nonumber \\ &\times& 
   \Theta ( 1  - 
 q_0 / \mu^\prime - z )   \ 
  \frac{\Delta_a(\mu ^{2}, \mu_0^2)}{\Delta_a(\mu ^{\prime 2}, \mu_0^2)} \  
  P^{R}_{ab}\left(z, \alpha_s\left((1-z)^2\mu^{\prime 2}\right)\right)\widetilde{f}_{b}\left(\frac{x}{z},\mu ^{\prime 2}\right)  \;\;.
\end{eqnarray} 
One can notice that this formula agrees with Catani-Marchesini-Webber (CMW) result (see eqs.~(42), (49)  
and  Sec.~3.4  of~\cite{Marchesini:1987cf}).  In Ref.~\cite{Marchesini:1987cf}  this   equation was studied at LO with one-loop splitting  
kernels and running coupling. In PB  method it 
is   studied  at NLO with two-loop splitting kernels and running coupling.

\section{Mapping evolution scales to transverse momenta} 
Exploiting the angular ordering relation $q_{\bot}=(1-z)\mu^{\prime}$, eq.~(\ref{PBangular}) can be rewritten by changing the integration variable from the branching scale $\mu^{\prime}$ to the transverse momentum $q_{\bot}$. Depending on the $x$ value, two cases can be distinguished, $x  \geq    1-{q_0}  /  {\mu_0}$, $  1- {q_0} / {\mu_0} > x > 0 $ which were studied in detail in  \cite{Hautmann:2019biw}. 
Here we concentrate only on the first case when  $1 > x \geq 1-{q_0} / {\mu_0}$ for which eq.~(\ref{PBangular}) can be written as 
\begin{eqnarray}
\label{PBangular_2term_3}
 \widetilde{f}_{a}(x,\mu ^{2}) &=& 
 \Delta_a(\mu ^{2} , \mu _{0} ^{2}) 
 \widetilde{f}_{a}(x,\mu _{0} ^{2})+  
  \sum _{b}  \int   \frac{\textrm{d}q_{\bot}^2}{q_{\bot}^2}
  \int_{x}^{1}\textrm{d}z \ 
 \Theta ( q_\perp^2 - q_0^2 ) \ \Theta ( \mu^2 
(1-x)^2  - q_\perp^2)   
  \nonumber \\ 
&\times&  \Theta ( 1  - q_\perp / \mu -z ) \ 
 \frac{\Delta_a(\mu ^{2}, \mu_0^2)}{\Delta_a\left(q_{\bot}^2 / (1-z)^2,\mu_0^2\right)}  \ 
  P^{R}_{ab}\left(z, \alpha_s\left(q_{\bot}^2\right)\right)\widetilde{f}_{b}\left(\frac{x}{z},\frac{q_{\bot} ^{2}}{(1-z)^2}\right) \; .  
\end{eqnarray}

\section{Multiple emissions vs single emission}

With this we can compare PB approach with the Kimber-Martin-Ryskin-Watt (KMRW)  \cite{Kimber:1999xc}
 angular ordered TMD evolution equation: 
\begin{eqnarray}
\label{eq:KMRWTMD}
\widetilde{D}_a(x,\mu ^{2}, q_{\bot}^2)&=&T_a(\mu ^{2}, q_{\bot} ^{ 2})\sum _{b} \int _{x}^{1-\frac{q_{\bot}}{{q_{\bot}+ \mu}}} dz P^{R}_{ab}\left(z, \alpha_s(q_{\bot} ^{ 2})\right)\widetilde{f}_{b}\left(\frac{x}{z}, q_{\bot}^{ 2}\right) \;\;  , 
\end{eqnarray}
where  the Sudakov form factor is  
$
T_a(\mu^2, q_{\bot}^2) = \exp\left[-\int_{q_{\bot}^2}^{\mu^2}\frac{\textrm{d}q_{\bot}^{\prime 2}}{q_{\bot}^{\prime 2}}\sum_b \int_0^{1-\frac{q_{\bot}}{{q_{\bot}+ \mu}}}\textrm{d}z z P_{ba}^{R}\left(z,\alpha_s\left(q_{\bot}^{\prime 2}\right)\right) \right]  
$. 
The collinear density   $\widetilde{f}_{a}(x,\mu ^{2}) $ 
obeys the evolution equation 
\begin{eqnarray}
\label{eq:KMR}
  \widetilde{f}_{a}(x,\mu ^{2}) &=& \widetilde{f}_{a}(x,\mu _{0} ^{2})T_a(\mu ^{2}, \mu_0^2)    \nonumber \\ &+& \int_{\mu_{0} ^{2}}^{\left(\frac{\mu({1-x})  }{x}\right)^2}  \frac{d q_{\bot} ^{ 2}}{q_{\bot} ^{ 2}}\left(T_a(\mu ^{2}, q_{\bot} ^{ 2})\sum _{b} \int _{x}^{1-\frac{q_{\bot}}{{q_{\bot}+ \mu}}} dz P^{R}_{ab}\left(z, \alpha_s(q_{\bot} ^{ 2})\right)\widetilde{f}_{b}\left(\frac{x}{z}, q_{\bot}^{ 2}\right) \right)   \; . 
\end{eqnarray}
To compare PB with KMRW, we neglect the difference between $q_0$ and $\mu_0$ since this distinction is not used in KMRW approach. \\

In contrast to PB method, where the transverse momentum is acquired in multiple branchings, the KMRW is a single emission approach: the evolution is performed in  one scale  up to $q_{\bot}$ and the second scale is generated in the last emission. These two approaches differ also in the scales at which the Sudakov form factors and parton densities $\widetilde{f}_b$ are evaluated: KMRW uses  just the transverse momentum whereas the PB uses transverse momentum rescaled by $(1-z)$. Moreover, the integration limits are different as well. The KMRW limits lead to the situation when the Sudakov form factor can be larger than $1$ losing its probabilistic interpretation and has to be frozen to some value (see e.g \cite{Golec-Biernat:2018hqo}). 

Interesting remark concerns the Sudakov form factors in these two methods  \cite{Hautmann:2019biw}. 
In the PB approach 
the Sudakov form factor 
fulfills the property  
\begin{equation}
\label{sudsud}
 \Delta_{a}(\mu ^{2}, \widetilde{\mu} ^{2})\Delta_{a}(\widetilde{\mu} ^{2}, \mu_0^2) = \Delta_{a}(\mu ^{2},\mu_0^2)  \; 
\end{equation}
for any evolution scale $\widetilde{\mu}$ which is a consequence of its  
 probabilistic interpretation. Also, after changing the integration variables from $\mu^{\prime}$ to $q_{\bot}$, this property remains valid. This is however not the case for KMRW Sudakov form factor.\\

These two approaches were compared numerically \cite{Hautmann:2019biw} using the  
 TMD sets MRW-CT10nlo \cite{Bury:2017jxo} which were obtained according to KMRW angular ordering prescription  and which can be found in TMDlib and TMDplotter \cite{Hautmann:2014kza}. The same starting distribution  CT10nlo~\cite{Lai:2010vv} was used for both of the approaches for the purpose of the comparison. PB TMD set was obtained with $q_0 = 1 \;\textrm{GeV}$.

\begin{figure}
\begin{minipage}{0.32\linewidth}
\vspace{1cm}
\centerline{
\includegraphics[width=0.99\linewidth]{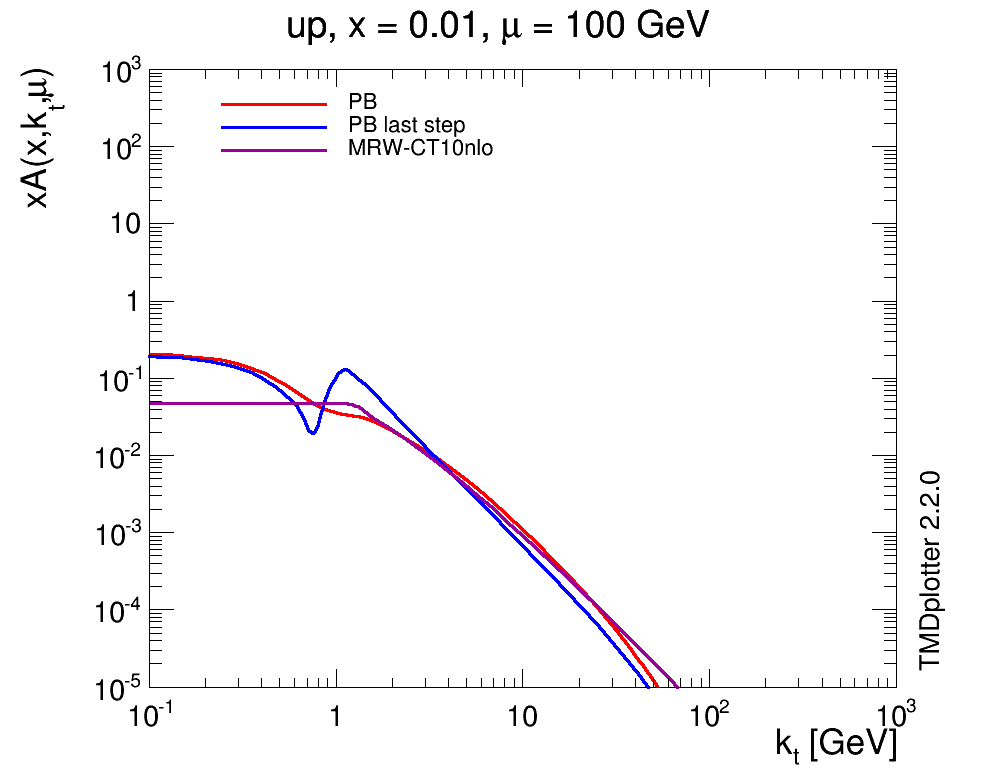}}
\caption[]{TMDs from PB and KMRW as functions of transverse momentum for up quark 
at  $x=0.01$ and  $\mu=100\;\textrm{GeV}$.   }
\label{fig:KMRWvsPB}
\end{minipage}
\hfill
\begin{minipage}{0.32\linewidth}
\centerline{\includegraphics[width=0.8\linewidth]{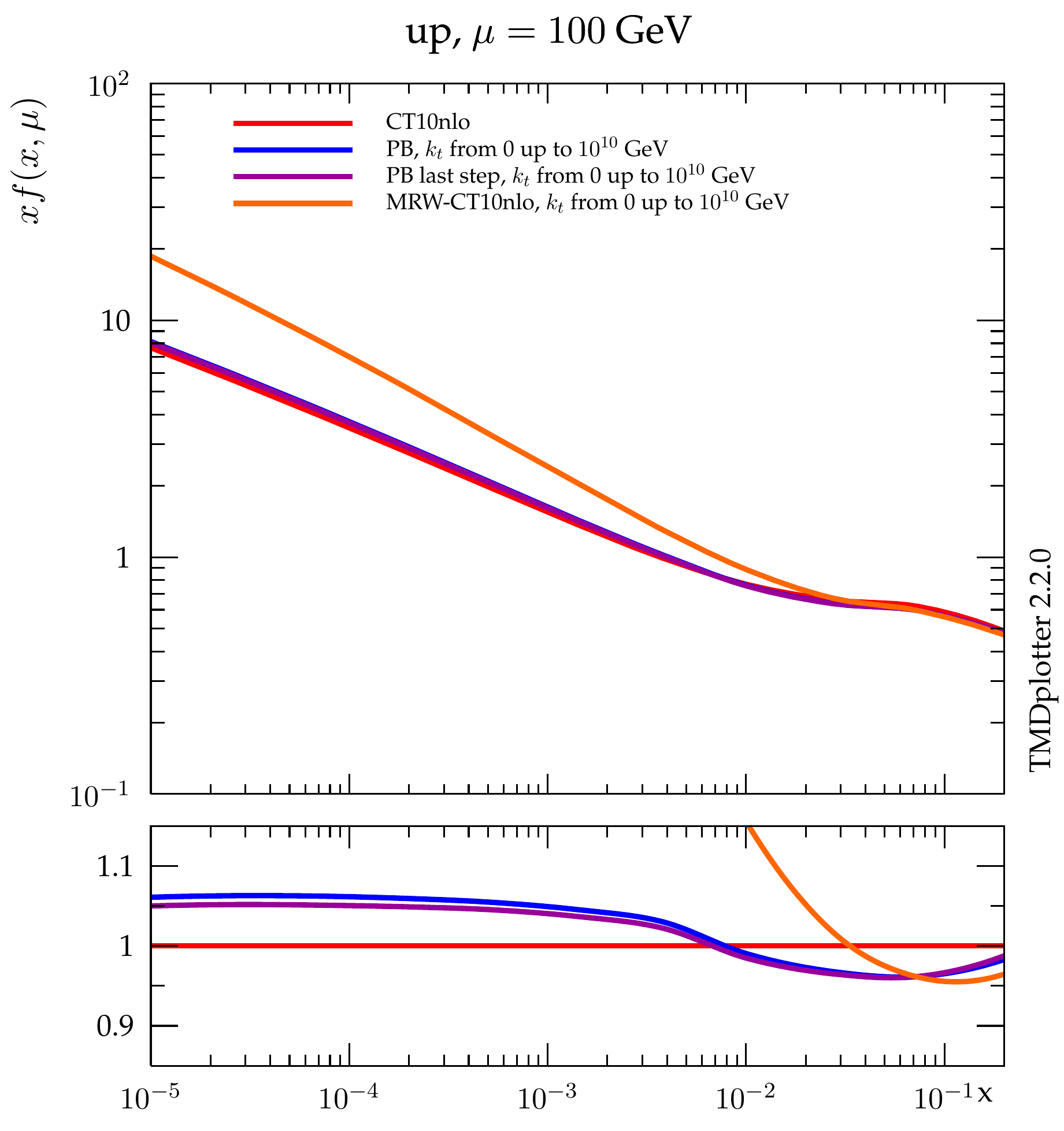}}
 \caption[]{The results of integrating TMDs  over all $ k_{\bot}$  as  functions of $x$.}
 \label{fig:KMRWvsPB_iTMDs}
\end{minipage}
\hfill
\begin{minipage}{0.32\linewidth}
\centerline{\includegraphics[width=0.95\linewidth]{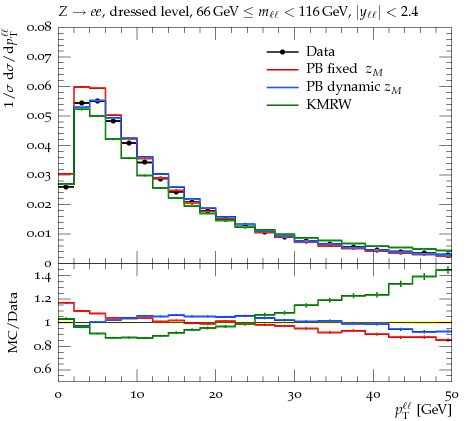}}
 \caption[]{ Predictions for the $Z$-boson $p_{\bot}$ spectrum obtained with the discussed TMDs.}
\label{fig:Zpt}
\end{minipage}
\hfill
\end{figure}

The result of the comparison is shown in fig.~\ref{fig:KMRWvsPB}. 
In the middle $k_{\bot}$ both approaches look very similar:
the differences between 
 the two  approaches  in the parton-density and  Sudakov-factor  and the phase space compensate for  KMRW not taking into account all previous emissions compared to PB. 
KMRW and PB differ in the low $k_{\bot}$ region where for KMRW we see the  intrinsic $k_{\bot}$ constant parametrization and  for PB the Gaussian intrinsic $k_{\bot}$   smeared during the evolution process. They differ also in the large $k_{\bot}$ region: KMRW has a  large $k_{\bot}$ tail coming from their treatment of the Sudakov form factor for $k_{\bot}> \mu$. 
To explore the effect of single vs multiple emission, {\it{PB last step}} curve was shown in fig.~\ref{fig:KMRWvsPB}. It was obtained with the same settings as the PB curve, with the only difference in the transverse momentum calculation: in {\it{PB last step}} the transverse momentum is generated only in the last emission. We can see that thanks to multiple emissions the unphysical jump in the distribution coming from matching of the intrinsic $k_{\bot}$ with the evolution is smeared. 

It is interesting to look also at the iTMDs shown in fig.~\ref{fig:KMRWvsPB_iTMDs}. 
In the lower part of the figure  the ratios of  iTMDs to  CT10nlo are plotted. 
None of the distributions integrate to CT10nlo, what is expected given that  the resolution scale $z_M$ is far from 1, and 
 the scale of the running coupling $\alpha_s$ is $q_{\bot}$. 
The high-$k_{\bot}$ tail of  MRW-CT10nlo results in 
 much higher integrated distribution than  other curves.

PB and  MRW-CT10nlo TMDs were used to obtain predictions for Z boson $p_{\bot}$ measurement at 8 TeV \cite{Aad:2015auj} shown in fig.~\ref{fig:Zpt}. We followed the procedure from \cite{Martinez:2018jxt}. We plotted also the prediction obtained with PB TMD Set-2 \cite{Martinez:2018jxt}, obtained with fixed $z_M$. Since for PB with dynamical $z_M$ and KMRW there is no uncertainty band available yet, only the central value for PB TMD Set-2 is shown as well. From the fig.~\ref{fig:Zpt} one can clearly see the improvement coming from dynamical resolution scale. MRW-CT10nlo does not describe  neither the  high $p_{\bot}$ region nor the low  $p_{\bot}$ slope. 

\section{Conclusions}
In this work we discussed the PB TMD evolution equation with angular ordering condition including dynamical soft gluon resolution scale $z_M$. Thanks to the dynamical $z_M$, we were able to establish connection of PB with other well known approaches, CMW and KMRS. 
We discussed the impact  of the dynamical soft gluon resolution scale on the precise predictions of Z boson $p_{\bot}$ and we showed improvement compared to using fixed $z_M$ value. 

\textbf{Acknowledgments} The results presented in this article were obtained in collaboration with Francesco Hautmann,   Aron Mees van Kampen and  Lissa Keersmaekers. We thank Hannes Jung for many discussions.

\vspace{-0.5cm}

\end{document}